\documentclass[twocolumn]{aastex63}

\usepackage{rotating}
\usepackage{acronym}
\usepackage{xspace}
\usepackage{mathtools}
\usepackage{enumitem}
\newcommand{\LIGOlabMIT}{\affiliation{LIGO Laboratory, Massachusetts Institute of Technology, 185 Albany Street, Cambridge, MA 02139, USA}}
\newcommand{\MKI}{\affiliation{Department of Physics and Kavli Institute for Astrophysics and Space Research, Massachusetts Institute of Technology, 77 Massachusetts Avenue, Cambridge, MA 02139, USA}}
\newcommand{\CIERA}{\affiliation{Center for Interdisciplinary Exploration and Research in Astrophysics (CIERA) and Department of Physics and Astronomy, Northwestern University, 1800 Sherman Avenue, Evanston, IL 60201, USA}}

\definecolor{chmagenta}{rgb}{0.54, 0.17, 0.88}

\graphicspath{{./}{figures/}}

\def\chieff{\ensuremath{\chi_\mathrm{eff}}\xspace}

\def\chip{\ensuremath{\chi_\mathrm{p}}\xspace}
\def\chione{\ensuremath{\chi_1}\xspace}
\def\chitwo{\ensuremath{\chi_2}\xspace}

\def\thetatwo{\ensuremath{\theta_2}\xspace}
\def\q{\ensuremath{q}\xspace}

\def\BF{\ensuremath{\mathcal{B}}\xspace}

\def\Phenom{\texttt{Phenom}\xspace}
\def\EOB{\texttt{EOB}\xspace}

\def\PrimaryMassApprox{\ensuremath{{\sim} 30\,M_\odot}\xspace}
\def\SecondaryMassApprox{\ensuremath{{\sim} 8\,M_\odot}\xspace}

\def\EffectiveSpin{\ensuremath{0.25^{+0.08}_{-0.11}}}
\def\PrimarySpinMag{\ensuremath{0.43^{+0.16}_{-0.26}}}
\def\PhenomAlow{\ensuremath{0.09}\xspace}
\def\PhenomAhigh{\ensuremath{0.60}\xspace}
\def\EOBAlow{\ensuremath{0.31}\xspace}
\def\EOBAhigh{\ensuremath{0.58}\xspace}
\def\CombinedAlow{\ensuremath{0.17}\xspace}
\def\CombinedAhigh{\ensuremath{0.59}\xspace}
\def\PhenomMassRatioLow{\ensuremath{0.25}\xspace}
\def\PhenomMassRatioHigh{\ensuremath{0.45}\xspace}
\def\EOBMassRatioLow{\ensuremath{0.21}\xspace}
\def\EOBMassRatioHigh{\ensuremath{0.31}\xspace}
\def\PhenomEffectiveSpinHigh{\ensuremath{0.29}\xspace}
\def\EOBEffectiveSpinHigh{\ensuremath{0.34}\xspace}

\def\BFModelBApprox{\ensuremath{1.6}\xspace}
\def\BFModelCApprox{\ensuremath{6}\xspace}
\def\BFModelCSimeq{\ensuremath{6.4}\xspace}

\def\BFModelEApprox{\ensuremath{2}\xspace}
\def\BFModelFApprox{\ensuremath{2}\xspace}
\def\BFModelGApprox{\ensuremath{350}\xspace}
\def\BFModelAEOBApprox{\ensuremath{0.97}\xspace}
\def\BFModelDEOBApprox{\ensuremath{20}\xspace}

\def\LikelihoodDifferencePhenomDex{\ensuremath{\lesssim\,1.5}\xspace}
\def\LikelihoodDifferenceEOBDex{\ensuremath{\simeq\,2.7}\xspace}
\def\LikelihoodDifferenceEOB{\ensuremath{\gtrsim\,500}\xspace}

\def\qNinetyNineAll{\ensuremath{0.57}\xspace}

\def\qPhenomMedianModelA{\ensuremath{0.32}\xspace}
\def\qPhenomMedianModelD{\ensuremath{0.39}\xspace}
\def\qEOBMedianModelA{\ensuremath{0.26}\xspace}
\def\qEOBMedianModelD{\ensuremath{0.39}\xspace}

\def\qLowEOBModelD{\ensuremath{0.34}\xspace}
\def\qHighEOBModelD{\ensuremath{0.47}\xspace}

\def\ChiEffLowPhenomModelA{\ensuremath{0.12}\xspace}
\def\ChiEffHighPhenomModelA{\ensuremath{0.30}\xspace}
\def\ChiEffLowPhenomModelD{\ensuremath{0.06}\xspace}
\def\ChiEffHighPhenomModelD{\ensuremath{0.22}\xspace}
\def\ChiEffLowEOBModelA{\ensuremath{0.19}\xspace}
\def\ChiEffHighEOBModelA{\ensuremath{0.36}\xspace}
\def\ChiEffLowEOBModelD{\ensuremath{0.10}\xspace}
\def\ChiEffHighEOBModelD{\ensuremath{0.23}\xspace}
\def\MaxChiEffApprox{\ensuremath{0.21}\xspace}

\def\ChiPNonspinningNinety{\ensuremath{0.28}\xspace}
\def\ChiPMedianModelE{\ensuremath{0.027}\xspace}

\def\ChiOneDifference{\ensuremath{0.03}\xspace}
\def\ChiTwoNinety{\ensuremath{0.50}\xspace}

\def\ChiTwoNinetyModelDEOB{\ensuremath{0.62}\xspace}

\def\ThetaTwoNinetyModelC{\ensuremath{18.5^{\circ}\xspace}}
\def\ThetaTwoNinetyModelD{\ensuremath{18.1^{\circ}\xspace}}
\def\ThetaTwoNinetyModelDEOB{\ensuremath{16.3^{\circ}\xspace}}
\def\ThetaTwoNinetyModelE{\ensuremath{3.2^{\circ}\xspace}}

\def\SecondarySpinMandel{\ensuremath{0.64}\xspace}

\def\SDRatioChiOne{\ensuremath{5.5}\xspace}

\acrodef{GW}{gravitational-wave}
\acrodef{LIGO}{Laser Interferometer Gravitational-Wave Observatory}
\acrodef{BH}{black hole}
\acrodef{BBH}{binary black hole}
\acrodef{NS}{black hole}
\acrodef{BNS}{binary neutron star}
\acrodef{O1}{first observing run}
\acrodef{O2}{second observing run}
\acrodef{O3}{third observing run}
\acrodef{O3a}{first half of the third observing run}
\acrodef{SNR}{signal-to-noise ratio}
\acrodef{GR}{general relativity}
\acrodef{PSD}{power spectral density}
\acrodef{AGN}{active galactic nucleus}
\acrodef{LVC}{LIGO Scientific \& Virgo Collaboration}
\acrodef{HM}{higher-order multipole}

\shorttitle{Interpreting the Spins of GW190412}
\shortauthors{Zevin et al.}

\begin{document}

\title{You Can't Always Get What You Want:\\The Impact of Prior Assumptions on Interpreting GW190412}

\author[0000-0002-0147-0835]{Michael Zevin}\thanks{zevin@u.northwestern.edu}
\CIERA

\author[0000-0003-3870-7215]{Christopher P.~L. Berry}
\CIERA
\affiliation{SUPA, School of Physics and Astronomy, University of Glasgow, Glasgow G12 8QQ, UK}

\author[0000-0002-0403-4211]{Scott Coughlin}
\CIERA

\author[0000-0002-5833-413X]{Katerina Chatziioannou}
\affiliation{Center for Computational Astrophysics, Flatiron Institute, 162 5th Avenue, New York, NY 10010, USA}

\author[0000-0003-2700-0767]{Salvatore Vitale} \LIGOlabMIT \MKI

\begin{abstract}

GW190412 is the first observation of a black hole binary with definitively unequal masses. 
GW190412's mass asymmetry, along with the measured positive effective inspiral spin, allowed for inference of a component black hole spin: the primary black hole in the system was found to have a dimensionless spin magnitude between \CombinedAlow and \CombinedAhigh ($90\%$ credible range). 
We investigate how the choice of priors for the spin magnitudes and tilts of the component black holes affect the robustness of parameter estimates for GW190412, and report Bayes factors across a suite of prior assumptions. 
Depending on the waveform family used to describe the signal, we find either marginal to moderate (\BFModelFApprox:$1$--\BFModelCApprox:$1$) or strong ($\gtrsim$\,\BFModelDEOBApprox:$1$) support for the primary black hole being spinning compared to cases where only the secondary is allowed to have spin. 
We show how these choices influence parameter estimates, and find the asymmetric masses and positive effective inspiral spin of GW190412 to be qualitatively, but not quantitatively, robust to prior assumptions. 
Our results highlight the importance of both considering astrophysically motivated or population-based priors in interpreting observations and considering their relative support from the data. 

\end{abstract}

\section{Introduction}\label{sec:intro}

GW190412~\citep{GW190412} was the first reported observation of a \ac{BBH} from the \ac{O3} of the Advanced LIGO~\citep{aLIGO} and Advanced Virgo~\citep{aVirgo} detector network. 
GW190412's source is the first system to have definitively unequal masses~\citep[see][]{GWTC1}, with the primary \ac{BH} being \PrimaryMassApprox and the secondary \ac{BH} being \SecondaryMassApprox. 
In addition to unveiling emission from \acp{HM}, this asymmetry allowed for enhanced constraints on the intrinsic and extrinsic parameters of the \ac{BBH} system. 

The spins of compact binary components are difficult to measure from \ac{GW} signals~\citep{Poisson1995,Vitale2014,GW150914_properties,Purrer2016}. 
Typically, spin constraints are presented in terms of mass-weighted combinations of the two component spins:
the effective inspiral spin
\begin{equation}\label{eq:chi_eff}
\chieff = \frac{m_1 \chi_1 \cos\theta_1 + m_2 \chi_2 \cos\theta_2}{m_1 + m_2},
\end{equation}
where $m_1\,\geq\,m_2$ are the component masses, $\chi_i$ are the dimensionless spin magnitudes, and $\theta_{i}$ are the angles between the spins and the Newtonian orbital angular momentum, $\vec{L}$, encodes information about the spin components aligned with the orbital angular momentum~\citep{Damour2001,Racine2008,Santamaria2010,Ajith2011}, whereas in-plane spins are characterized by the effective precession spin~\citep[][]{Schmidt2015}
\begin{equation}
\chip = \max\left\{\chione \sin \theta_1, \frac{q(4q+3)}{4+3q} \chitwo \sin \theta_2\right\}. 
\end{equation}
The \ac{LVC} reported an effective spin for GW190412 of ${\chieff=\EffectiveSpin}$~\citep[median and $90\%$ credible interval;][]{GW190412}. 
Since \chieff is positive and constrained away from zero, at least one of the \acp{BH} in the GW190412 system had a spin direction in the same hemisphere as $\vec{L}$ during the \ac{GW} inspiral. 
GW190412 also exhibited marginal hints of orbital precession, which is consistent with at least one of the \ac{BH} spins being nonzero. 

A \ac{BBH} with $\chieff > 0$ has been observed before in GW151226~\citep[][]{GW151226,Miller2020}, and potentially in GW170729~\citep{GWTC1,Chatziioannou2019}. 
However, the larger mass of the primary \ac{BH} in GW190412 relative to the secondary \ac{BH} allowed for the spin of the primary to be inferred as ${\chione = \PrimarySpinMag}$. 
This is because when $m_1 \gg m_2$ the primary spin is much more important in determining the dynamics of the system (as illustrated by the mass weighting in \chieff and \chip), and we are less sensitive to the value of the secondary spin. 
GW190412 therefore is the first high-significance detection of a compact binary system with an observable component spin.\footnote{A potential \ac{BBH} with a highly spinning primary component was reported in \citet{Zackay2019a}, though the astrophysical origin of the signal is under debate~\citep{Ashton2020,Nitz2020}, and due to the low signal-to-noise of this event, the spin interpretation depends heavily on the choice of prior~\citep{Galaudage2019,Huang2020,Nitz2020}.} 

GW190412's primary spin may be difficult to reconcile with theoretical modeling of massive binary stars in isolation. 
Detailed modeling of core--envelope interaction in massive stars finds angular momentum transport to be highly efficient, driving stellar cores to extremely slow rotation prior to their collapse into a compact object~\citep{Kushnir2016,Zaldarriaga2018,Fuller2019a,Fuller2019b,Belczynski2020}. 
This theoretical underpinning is corroborated by current \ac{GW} catalogs, which contain systems that are mostly consistent with $\chieff \approx 0$~\citep{O2RandP,Miller2020,Nitz2020,Venumadhav2020}.
Though the birth spins of some \acp{BH} in high-mass X-ray binaries have been interpreted as near extremal ($\chi \approx 1$; see \citealt{Miller2015} and references therein), it is unclear whether these systems will evolve to be \acp{BBH} that merge within a Hubble time~\citep[e.g.,][]{Belczynski2012,Qin2019}. 
Following this reasoning, multiple groups have proposed that the high spin of the primary \ac{BH} in GW190412 is the result of an alternative formation scenario to canonical isolated binary evolution, such as dynamical assembly in young star clusters~\citep{DiCarlo2020}, hierarchical mergers in massive stellar clusters~\citep{Gerosa2020,Kimball2020,Rodriguez2020}, \ac{AGN} disks~\citep{Tagawa2020}, Population III stars~\citep{Kinugawa2020}, and mergers induced from the secular evolution in hierarchical systems~\citep{Hamers2020}. 

On the other hand, the second-born \ac{BH} in \ac{BBH} merger progenitors can be significantly spun up through tidal locking of the stellar core with the first-born \ac{BH}~\citep{Qin2018,Bavera2020}. 
If GW190412 could instead be explained by a highly spinning secondary \ac{BH}, the standard isolated formation scenario with a low-spinning primary could again be viable. 
To this end, \citet{Mandel2020} provide a reinterpretation of the \ac{LVC} analysis~\citep{GW190412} using a prior motivated by theoretical predictions of \ac{BBH} progenitors formed in isolation. 
Assuming a prior with a zero-spin primary \ac{BH} and a secondary \ac{BH} whose spin projection is aligned with the orbital angular momentum, \citet{Mandel2020} reweight the public posterior samples of GW190412~\citep{GW190412datarelease}, effectively interpreting the measured value of $\chieff$ as originating from the secondary's spin rather than the primary's.  
To compensate for the nonzero effective spin of GW190412, the reweighted posteriors from this analysis point to a highly spinning secondary \ac{BH} with $\chi_2 \gtrsim \SecondarySpinMandel$~\citep{Mandel2020}. 

Though predictions for the formation rate of these systems are highly sensitive to the uncertain prescription for natal \ac{BH} spins, recent work has found that for systems with asymmetric masses such as GW190412, the highly spinning secondary \ac{BH} interpretation is more probable from an isolated evolution standpoint than a moderately spinning primary~\citep[e.g.,][]{Olejak2020}. 
This is consistent with the current catalog of \acp{GW}, since individual spins are poorly constrained in all previously observed \acp{BBH}~\citep{O2RandP}. 
However, even this formation mechanism struggles to accommodate GW190412, as systems where the secondary \ac{BH} has been significantly spun up due to tidal interactions have short merger timescales and a merger rate in the local universe that is at least an order of magnitude lower than what is estimated for GW190412-like systems~\citep{Safarzadeh2020c}. 
 
Nonetheless, while various assumptions may be made to represent the prior belief for parameters given an astrophysical model, it is critical to determine whether a given model is supported by the data. 
The amount by which the data supports a specific model (in this work, a prior) is encoded in the Bayesian evidence. 
While varying prior assumptions will yield differing parameter estimates, the ratio of evidences between models---the \emph{Bayes factor} \BF---indicates whether any one prior assumption is favored or disfavored by the data compared to another. 
This is particularly important to verify for the case of strong priors, since they might drive the posteriors to potentially arbitrary values at the expense of the evidence: \emph{if you torture the data long enough, it will confess to anything}~\citep{Coase1982}. 
For example, in the analysis of GW151226~\citep{GW151226}, \citet{Vitale2017b} showed how if one uses a prior that enforces small ($\sim 0.1$) spin magnitudes, the evidence decreases by a factor of $50$ compared to a uniform prior, while the posteriors still look reasonable. 
It is only by comparing evidences between models, i.e.\ calculating Bayes factors, that one can assess which model is better described by the data. 

In this Letter, we explore various prior assumptions for the interpretation of GW190412 and calculate Bayes factors between these model assumptions. 
The priors we choose are motivated by various astrophysical models presented in the literature, with a particular focus on the spin of the second-born \ac{BH}, and the astrophysically relevant question of whether the primary is spinning.  

In Section~\ref{sec:analysis} we explain the various prior assumptions we choose when analyzing the data, and their astrophysical motivation. 
We present Bayes factors across these prior assumptions in Section~\ref{sec:BF}, and examine the impact of differing prior assumptions on the parameter estimation for GW190412 in Section~\ref{sec:PE}. 
In Section~\ref{sec:discussion} we discuss the results of our analysis and their impact on the interpretation of GW190412, and comment on astrophysical implications.

\section{Data Analysis and Prior Assumptions}\label{sec:analysis}

To investigate the impact of prior assumptions on the inferred parameters of GW190412 and the Bayes factors between these assumptions, we perform parameter estimation using a suite of prior assumptions motivated by various astrophysical predictions. 
We use the publicly available data for GW190412~\citep{GW190412datarelease} and follow the parameter-estimation procedure used in \citet{GW190412}.
Our results are produced using a highly parallelized version of \textsc{Bilby}~\citep{Ashton2019,Smith2019,Romero-Shaw2020a}, which computes posterior probability distributions for the properties of the source as well as model evidences. 

We use both the \Phenom and \EOB families of waveform approximants in our analysis.\footnote{Different waveform models are referred to as ``approximants'' throughout to differentiate between waveform approximant models and the models representing different prior configurations.} 
We use \texttt{IMRPhenomPv3HM}~\citep{Khan2019,Khan2020} and \texttt{SEOBNRv4PHM}~\citep{Pan2014,Babak2017,Ossokine2020}, both of which include the effects of spin precession and \ac{HM} moments. 
Inclusion of \acp{HM} in waveform approximants is crucial for the parameter estimation of GW190412, as this more complete physical picture of the \ac{GW} signal is necessary to accurately constrain the mass ratio (${q = m_2/m_1 < 1}$) and spins~\citep{VanDenBroeck2007,Graff2015,CalderonBustillo2016,Varma2017,GW190412}.  
Systematic differences are expected between analyses using \Phenom and \EOB approximants, as evident in \cite{GW190412}. 
Though we use the \Phenom approximant for all seven prior configurations described below, due to the computational cost of the \EOB approximant we only run this with two exemplary prior configurations. 

The priors we consider are: 
\begin{enumerate}[label=\Alph*)]
    \item Uniform in spin magnitude for both components, isotropic and unconstrained in spin tilts. 
    This uninformative prior is used in \citet{GW190412}; it does not make strong assumptions about spin orientations or magnitudes, and its broad support enables reweighting by different priors~\citep[e.g.,][]{Mandel2019,Thrane2019}.
    
    \item Uniform in spin magnitude and isotropic in spin tilt for the primary \ac{BH}, with a non-spinning secondary. 
    A spinning primary and a non-spinning secondary may be expected if \acp{BH} are born with small spins, but the larger \ac{BH} is the result of a previous \ac{BH} merger and has gone on to form a new binary in a dense stellar environment such as a globular or nuclear cluster~\citep{Fishbach2017,Gerosa2017a,Rodriguez2019,Gerosa2020,Kimball2020}. 
    In this scenario, we would typically expect the primary spin magnitude to be $\chione \sim 0.67$.
    
    \item Non-spinning primary \ac{BH} with unconstrained spin for the secondary \ac{BH}. 
    This is representative of an isolated formation scenario, with a secondary that can be spun up through tidal interactions~\citep{Qin2019,Bavera2020}. 
    The unconstrained spin tilt, however, allows for significantly misaligned spins, which are difficult to attain for \acp{BBH} in the standard isolated evolution scenario~\citep[e.g.,][]{Kalogera2000,Fryer2012,Rodriguez2016}. 
    
    \item Same as Model C, but with spin tilts constrained to be in the same hemisphere as the orbital angular momentum: ${\vec{\chi}_2 \cdot \vec{L} \geq 0}$ (${0^{\circ} \leq \theta_2 \leq 90^{\circ}}$). 
    This is similar to the prior assumption used in \cite{Mandel2020}. 
    
    \item Same as Model C, but with spin tilts for the secondary constrained to ${0^{\circ} \geq \theta_2 \geq 10^{\circ}}$. 
    This model has been used to represent near-aligned spins~\citep[e.g.,][]{Vitale2017a}, as predicted from the coevolution of isolated binaries and weak \ac{BH} natal kicks at birth. 
    
    \item Same as Model C, but with spin tilts for the secondary perfectly aligned with the orbital angular momentum (${\theta_2 = 0^{\circ}}$). 
    
    \item Non-spinning primary and secondary. 
    This is an extreme assumption that we expect will struggle to match the data due to the positive measured \chieff and marginal precessional information. 

\end{enumerate}
These configurations are summarized on the left side of Table~\ref{tab:table}. 
For all other parameters, we use priors analogous to those used by the \ac{LVC} in the analysis of GW190412~\citep{GW190412}. 

\begin{deluxetable*}{@{\extracolsep{4pt}}c  c c c c  c c c c}
\label{tab:table}
\tablecaption{\emph{Left:} Prior assumptions for component spin magnitudes and spin tilts in each model. 
We use short-hand for the distributions we consider: $\delta$ uses a fixed value, U denotes a uniform distribution, and ISO an isotropic distribution (uniform in $\cos(\theta))$. 
Angular assumptions are omitted when the spin magnitudes for that component are forced to zero. 
\emph{Right:} The maximum value of the log likelihood ($\log_{10}\mathcal{L}_{\rm max}$) and Bayes factors (\BF) for each model. 
Each Bayes factor is calculated relative to the uninformative \ac{LVC} prior for the respective waveform approximant (Model A and Model A-\EOB for \Phenom and \EOB, respectively). 
For reference, $\BF < 1$ ($\log_{10}\BF < 0$) means that data prefers the reference model, $\BF \gtrsim 3$:$1$ ($\log_{10} \BF \gtrsim 0.5$) indicates moderate evidence for the new hypothesis, and $\BF \gtrsim 10$:$1$ ($\log_{10} \BF \gtrsim 1.0$) indicates strong evidence for the new hypothesis. 
The rightmost column gives the estimated $1\,\sigma$ uncertainty in $\log_{10} \BF$. }
\tablehead{
\colhead{} & \multicolumn{4}{c}{Prior Assumption} & \multicolumn{4}{c}{Evidence} \\
\cline{2-5} \cline{6-9}
\colhead{Model} & \colhead{$\chi_1$} & \colhead{$\theta_1$} & \colhead{$\chi_2$} & \colhead{$\theta_2$} & \colhead{$\log_{10} \mathcal{L}_{\rm max}$} & \colhead{\BF} & \colhead{${\log_{10}}(\BF)$} &\colhead{$\sigma_{\log_{10}(\BF)}$}
}
\startdata
$\mathrm{A}$ & $\mathrm{U}[0,0.99]$ & $\mathrm{ISO}[0^\circ,180^\circ]$ & $\mathrm{U}[0,0.99]$ & $\mathrm{ISO}[0^\circ,180^\circ]$ & $77.0$ & $1.0$ & $0.00$ & $0.08$ \\
    $\mathrm{A}$-$\texttt{EOB}$ & $\mathrm{U}[0,0.99]$ & $\mathrm{ISO}[0^\circ,180^\circ]$ & $\mathrm{U}[0,0.99]$ & $\mathrm{ISO}[0^\circ,180^\circ]$ & $77.1$ & $1.0$ & $0.00$ & $0.10$ \\
    $\mathrm{B}$ & $\mathrm{U}[0,0.99]$ & $\mathrm{ISO}[0^\circ,180^\circ]$ & $\delta(0)$ & $-$ & $76.5$ & $6.2\,\times\,10^{-1}$ & $-0.20$ & $0.09$ \\
    $\mathrm{C}$ & $\delta(0)$ & $-$ & $\mathrm{U}[0,0.99]$ & $\mathrm{ISO}[0^\circ,180^\circ]$ & $75.6$ & $1.5\,\times\,10^{-1}$ & $-0.80$ & $0.09$ \\
    $\mathrm{D}$ & $\delta(0)$ & $-$ & $\mathrm{U}[0,0.99]$ & $\mathrm{ISO}[0^\circ,90^\circ]$ & $75.5$ & $3.6\,\times\,10^{-1}$ & $-0.44$ & $0.09$ \\
    $\mathrm{D}$-$\texttt{EOB}$ & $\delta(0)$ & $-$ & $\mathrm{U}[0,0.99]$ & $\mathrm{ISO}[0^\circ,90^\circ]$ & $74.4$ & $4.9\,\times\,10^{-2}$ & $-1.30$ & $0.10$ \\
    $\mathrm{E}$ & $\delta(0)$ & $-$ & $\mathrm{U}[0,0.99]$ & $\mathrm{ISO}[0^\circ,10^\circ]$ & $75.3$ & $5.3\,\times\,10^{-1}$ & $-0.27$ & $0.09$ \\
    $\mathrm{F}$ & $\delta(0)$ & $-$ & $\mathrm{U}[0,0.99]$ & $\delta(0^\circ)$ & $75.2$ & $5.0\,\times\,10^{-1}$ & $-0.30$ & $0.09$ \\
    $\mathrm{G}$ & $\delta(0)$ & $-$ & $\delta(0)$ & $-$ & $71.5$ & $2.8\,\times\,10^{-3}$ & $-2.54$ & $0.08$ \\
    \enddata
\end{deluxetable*}

\section{Bayes Factors}\label{sec:BF}

Given the observation of GW190412, we can identify which astrophysical model is best supported by the data. 
To quantify the relative support for different models, we would ideally use the \emph{odds ratio}; the odds ratio between models $\mathcal{M}_i$ and $\mathcal{M}_j$ is defined as
\begin{equation}
    \mathcal{O}_{\mathcal{M}_i,\mathcal{M}_j} = \frac{p(\mathcal{M}_i|d)}{p(\mathcal{M}_j|d)},
\end{equation}
where $p(\mathcal{M}_i|d)$ is the posterior probability of model $\mathcal{M}_i$ given the data $d$. 
Using Bayes's theorem, the odds ratio can be written as
\begin{equation}
    \mathcal{O}_{\mathcal{M}_i,\mathcal{M}_j} = \frac{p(\mathcal{M}_i)}{p(\mathcal{M}_j)}\frac{p(d|\mathcal{M}_i)}{p(d|\mathcal{M}_j)}.
    \label{eq:odds}
\end{equation}
Here the first term is the prior odds: our expectation for the relative probabilities of the two models before observing the data. 
For example, predictions for the local \ac{BBH} merger rate from isolated binary evolution range from $\sim\,8$--$200~{\mathrm{Gpc}^{-3}\,\mathrm{yr}^{-1}}$~\citep[e.g.,][]{Eldridge2017,Klencki2018a,Giacobbo2018b,Giacobbo2020} while predictions for the local \ac{BBH} merger rate through dynamical assembly in globular clusters range from $\sim 0.8$--$35~{\mathrm{Gpc}^{-3}\,\mathrm{yr}^{-1}}$~\citep[e.g.,][]{Fragione2018,Hong2018,Rodriguez2018a}; thus, from the ratio of these predicted rates one may estimate a prior odds between the two channels of $\sim 0.2$--$250$. 
In addition to considering expected rates, prior odds could also factor in our belief in the accuracy of different physical prescriptions, for example, the efficiency of angular momentum transport in massive stars.
Given the uncertainties in the prior odds, we concentrate on the second term in Equation~\eqref{eq:odds}, the Bayes factor: the ratio of evidences for the two models. 

The model evidence, or marginalized likelihood, is
\begin{equation}
    p(d|\mathcal{M}_i) = \int \mathrm{d}\boldsymbol{\vartheta}_i\, \mathcal{L}(\boldsymbol{\vartheta}_i) p(\boldsymbol{\vartheta}_i|\mathcal{M}_i),
\end{equation}
where the integral is over the parameters $\boldsymbol{\vartheta}_i$ describing our source (masses, spins, etc.), $\mathcal{L}(\boldsymbol{\vartheta}_i) = p(d|\boldsymbol{\vartheta}_i)$ is the likelihood of the parameters~\citep{Cutler1994}, and $p(\boldsymbol{\vartheta}_i|\mathcal{M}_i)$ is our prior probability density on the parameters within model $\mathcal{M}_i$, as described in Section~\ref{sec:analysis}. 
Thus, the Bayes factor is given by
\begin{equation}\label{eq:BF}
    \BF_{\mathcal{M}_i,\mathcal{M}_j} =
    \frac{\int \mathrm{d}\boldsymbol{\vartheta}_i\, p(d|\boldsymbol{\vartheta}_i) p(\boldsymbol{\vartheta}_i|\mathcal{M}_i)}
    {\int \mathrm{d}\boldsymbol{\vartheta}_j\, p(d|\boldsymbol{\vartheta}_j) p(\boldsymbol{\vartheta}_j|\mathcal{M}_j)}.
\end{equation}

When considering models with more parameters, or with parameters allowed to vary on a larger domain, we expect that we may be able to fit the data better, giving higher likelihoods. 
In calculating evidences, this is counterbalanced by the increased prior volume: as we spread the total prior probability (which must integrate to $1$) over a larger volume where the likelihood can have potentially negligible support, its density around the maximum likelihood region may decrease, resulting in a lower evidence. 
This \emph{Occam factor} allows the Bayes factor to be used to determine if more complicated models are needed to explain data~\citep[][Chapter 28]{MacKay2003}.

When considering spins measured with \ac{GW} observations, we are typically only sensitive to particular mass-weighted combinations of the $6$ spin degrees of freedom~\citep{Poisson1995,Chatziioannou2014,Purrer2016,Vitale2017b}. 
Therefore, it may be possible to fit the data well by assuming only a single component is spinning, and we would not anticipate a strong preference in favor of a more complicated model including two spinning bodies.  
In cases when there is a large asymmetry in masses the secondary spin may become irrelevant, and the properties of the signal may be completely determined by the primary spin. 
When the secondary spin has negligible impact on the likelihood, we expect there will be no preference between models with and without a secondary spin as it is unconstrained and its introduction incurs no Occam factor penalty.\footnote{Analogously, when spin precession is not measurable, such that the posterior distribution for \chip is identical to the prior, we expect no preference between using a waveform approximant that includes spin precession and one that only includes the effects of the spin components aligned with the orbital angular momentum, assuming that the priors on the aligned components of the spin are equivalent.}

In Table~\ref{tab:table} we show Bayes factors for each prior configuration compared to the standard \ac{LVC} prior (Model A). 
Bayes factors could also be used to compare waveform approximants; for example, the Bayes factor between the \Phenom and \EOB approximants using the \ac{LVC} prior (Models A and A-\EOB) is $\BFModelAEOBApprox$:$1$, indicating no preference for one of these approximants over the other. 
Since we focus on the impact of differing prior configurations, unless otherwise specified we only compare Bayes factors between parameter-estimation runs that use the same waveform approximant (e.g., the Bayes factor for Model D is calculated relative to Model A, and the Bayes factor of Model D-\EOB is calculated relative to Model A-\EOB). 
Model A is preferred over the other prior configurations; the extra freedom allowed by having two spinning bodies enables a better fit to the data (as illustrated by the maximum likelihood value), and this improvement is sufficient to overcome the Occam factor from the larger prior volume. 

Despite the significant asymmetry in masses, the secondary spin still has some impact on the signal, as can be seen by comparing the Bayes factor between Model A and Model B (\BFModelBApprox:$1$ with the \Phenom approximant). 

We find marginal to moderate support for Model A relative to prior configurations where only the secondary is spinning (Models C--F) with the \Phenom approximant, and strong support for Model A relative to non-spinning primary configurations when using the \EOB approximant. 
For the $\chione = 0$ configurations and the \Phenom approximant, we find the greatest support for Model E, which is only disfavored relative to our fiducial prior configuration by a Bayes factor of ${\simeq \BFModelEApprox}$:$1$. 
As the opening angle for $\theta_2$ increases, we see a decreasing trend in the Bayes factor that is likely due to the Occam factor suffered by the prior configurations with larger possible misalignment, since the maximum likelihood across these three models is relatively constant. 
With a non-spinning primary, the secondary \ac{BH} needs to have significant spin aligned with the orbital angular momentum in order to match the observed signal. 
Therefore, the tilt is constrained to be small, and there is little in-plane spin. 
Though precession is possible in Models C--E, it is not possible to have a large \chip given both the mass ratio and the need to match the \chieff measurement.
With the \Phenom approximant, the case where we can draw the most confident conclusion is in comparison to the prior configuration with zero spins, which is disfavored by a Bayes factor of $\gtrsim\,\BFModelGApprox$:$1$.

We find strong support against the non-spinning primary hypothesis when using the \EOB approximant. 
The maximum likelihood value using the \ac{LVC} prior is greater than that of the non-spinning primary, aligned-spin secondary prior used in \cite{Mandel2020} by a factor of \LikelihoodDifferenceEOB.  
Though the \ac{LVC} prior configuration has a larger prior volume, the strong support in the data for a spinning primary leads to a Bayes factor of $\gtrsim\,\BFModelDEOBApprox$:$1$ relative to the non-spinning primary hypothesis. 
We discuss implications of these Bayes factors further in Section~\ref{sec:discussion}.

\section{Parameter Estimation}\label{sec:PE}

Prior assumptions can have a strong effect on the measurement of intrinsic and extrinsic parameters inherent to a \ac{BBH} coalescence. 
Here, we investigate the robustness of parameter estimates for GW190412 across our various prior assumptions, with a particular focus on spin parameters. 

\pagebreak
\subsection{Mass Ratio}

GW190412 is the first \ac{BBH} with definitely unequal masses, with a reported mass ratio at the $90\%$ credible level of ${\PhenomMassRatioLow \leq q \leq \PhenomMassRatioHigh}$ using the \Phenom approximant and ${\EOBMassRatioLow \leq q \leq \EOBMassRatioHigh}$ using the \EOB approximant~\citep{GW190412}. 
In Figure~\ref{fig:corner} we show the posterior distributions for \q across our different priors and waveform approximants. 
Aside from the (strongly disfavored) Model G, which does not allow for spins in either \ac{BH}, we find the mass ratio to be constrained to ${q \lesssim \qNinetyNineAll}$ at the $99\%$ credible level. 

There is a noticeable difference in the posterior distribution for \q when using priors where the primary is spinning compared to those where only the secondary is spinning. 
We find that the posterior for \q pushes to larger values when $\chione = 0$, with a median of \qPhenomMedianModelD (\qEOBMedianModelD) in Model D compared to \qPhenomMedianModelA (\qEOBMedianModelA) in Model A when using the \Phenom (\EOB) approximant. 
This change in \q results in a more massive secondary that can more easily account for the observed effective spins.

\subsection{Aligned Spin and Precession}

\begin{figure}[t!]
    \centering
    \includegraphics[width=\columnwidth]{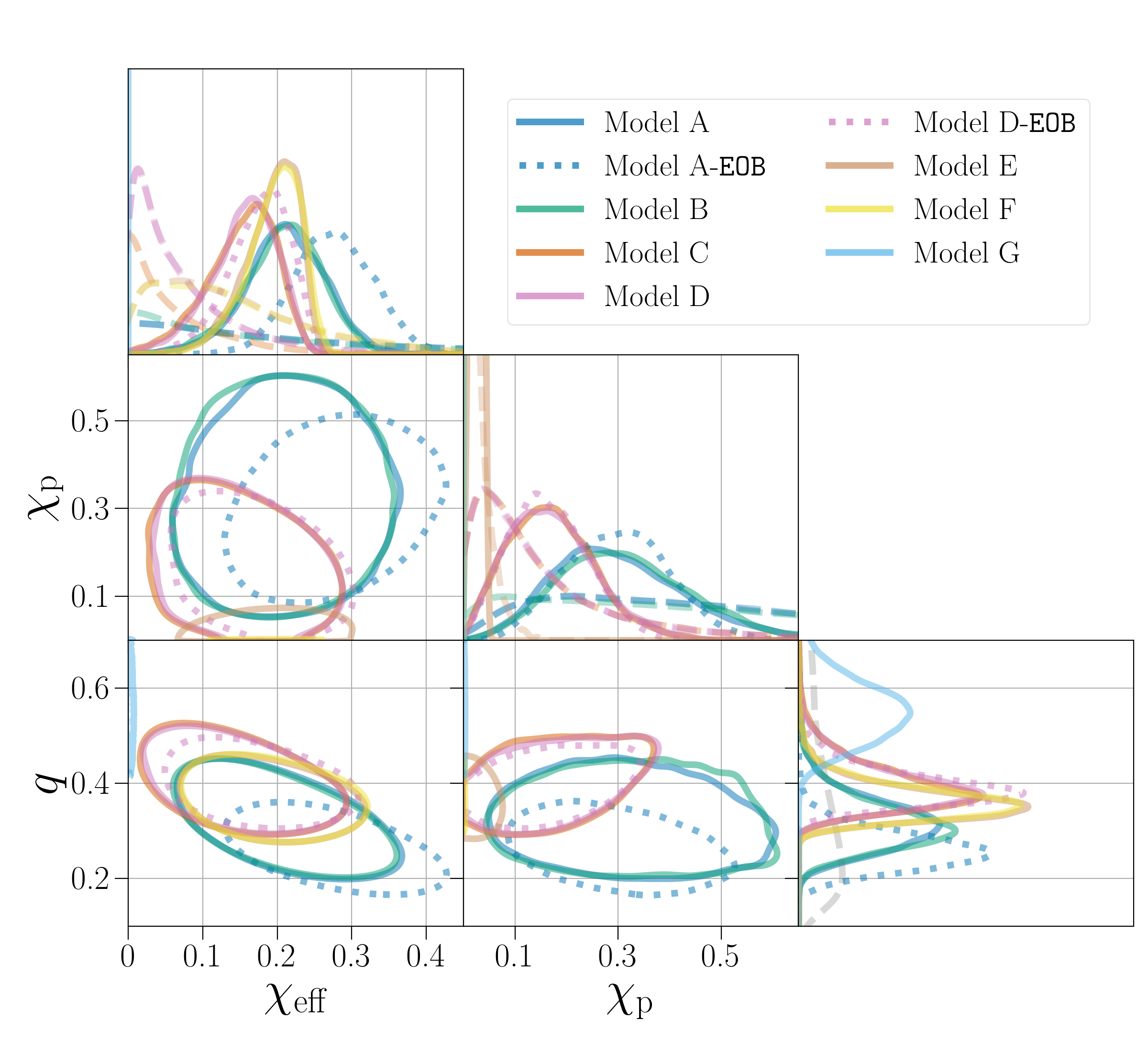}
    \vspace{-0.8cm}
    \caption{Joint and marginalized posterior distributions for the effective inspiral spin \chieff, effective precession spin \chip, and mass ratio \q. 
    The posteriors recovered for each prior configuration are shown using different colors. 
    Posteriors using the \Phenom approximant are shown in solid lines, and the subset of posteriors using the \EOB approximant are shown in dotted lines. 
    In the panels showing marginalized posterior distributions, the prior distributions for each configuration are shown with corresponding dashed lines; for \q the prior distribution is the same for all configurations and we display it with a single gray dashed line. 
    We see mild, yet noticeable differences in the posterior distributions for \q and \chieff when we constrain the primary spin to $\chione=0$, though we still recover asymmetric masses and a nonzero effective spin at high confidence for all runs with reasonable Bayes factors.
    For the non-spinning configuration (Model G) only $\chieff=0$ and $\chip=0$ are allowed. 
    }
    \label{fig:corner}
\end{figure}

When the primary is forced to be non-spinning, the effective inspiral spin migrates to lower values (lower left panel of Figure~\ref{fig:corner}); as apparent in Equation~\eqref{eq:chi_eff}, when $\chione = 0$, $|\chieff| \leq m2/(m1+m2)$. 
Using the \Phenom (\EOB) approximant, we find \chieff to be \ChiEffLowPhenomModelD--\ChiEffHighPhenomModelD (\ChiEffLowEOBModelD--\ChiEffHighEOBModelD) for the non-spinning primary Model D compared to \ChiEffLowPhenomModelA--\ChiEffHighPhenomModelA (\ChiEffLowEOBModelA--\ChiEffHighEOBModelA) for Model A at the $90\%$ credible level. 
Using the source parameters derived with the \ac{LVC} prior, the \emph{largest} \chieff that can be attained from a system with a non-spinning primary of mass ${m_1\,\PrimaryMassApprox}$ and a spinning secondary with mass ${m_2\,\SecondaryMassApprox}$ is ${\chieff|_{\chione=0} \lesssim \MaxChiEffApprox}$. 
Thus, prior configurations with a non-spinning primary need to compensate by jointly increasing \q and decreasing \chieff. 
However, for all our prior configurations where at least one \ac{BH} is spinning we find ${\chieff\,>\,0.08}$ at the 90\% credible level. 

Considering in-plane spins, \chip shows a larger variation between the prior configurations. 
This is expected, since \chip affects the likelihood only mildly and our prior configurations put restrictions on spin tilts. 
For both waveform approximants, when the primary is non-spinning \chip is ${\lesssim \ChiPNonspinningNinety}$ at the $90\%$ level, and rails against the physical boundary of $\chip=0$ (consistent with no precession). 
The median posterior value for \chip drops even more precipitously when a large degree of misalignment is not allowed; for Model E we recover a median \chip of \ChiPMedianModelE. 
Thus, if indeed the primary \ac{BH} is non-spinning, the marginal hints of precession in GW190412 disappear and the system is consistent with having a perfectly aligned secondary spin.

\subsection{Component Spins}

\begin{figure}[t!]
    \centering
    \includegraphics[width=\columnwidth]{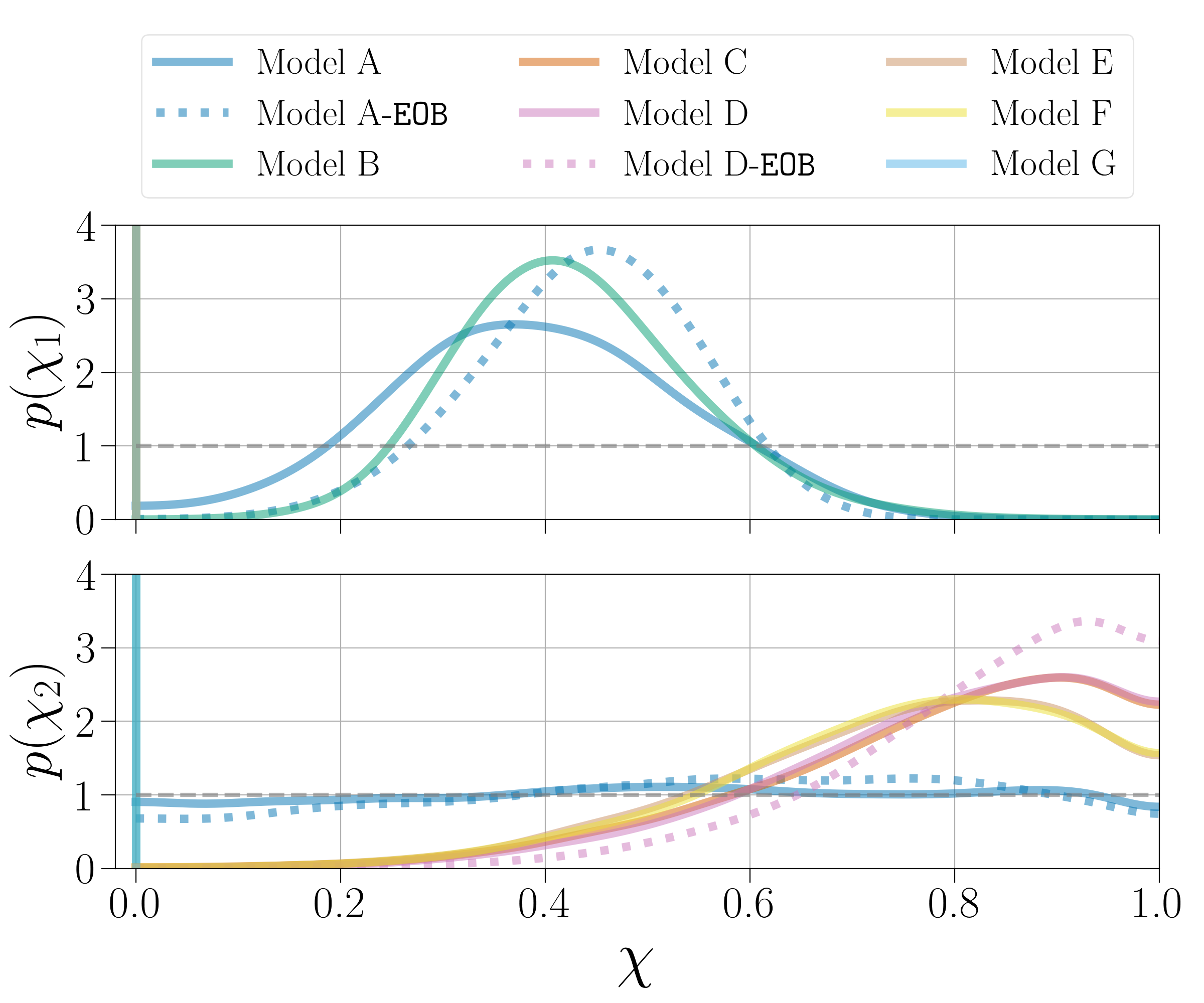}
    \caption{Marginalized spin distributions for the primary dimensionless \ac{BH} spin (\chione; \emph{top}) and the secondary dimensionless \ac{BH} spin (\chitwo; \emph{bottom}). 
    Posteriors recovered for each prior configuration are in different colors, and the flat priors for each spin magnitude are shown as a gray dashed line. 
    Posteriors attained using the \Phenom approximant are shown in solid lines, and the subset of posteriors from the \EOB approximant are shown in dotted lines. 
    When the primary \ac{BH} is non-spinning, \chitwo is constrained to higher values and consistent with maximally spinning ($\chitwo = 1$). 
    For priors where the primary or secondary \ac{BH} are forced to be non-spinning, the posterior is shown by a vertical line at $\chi=0$. 
    }
    \label{fig:a1_a2}
\end{figure}

In Figure~\ref{fig:a1_a2} we show marginalized posterior distributions for the two component spins, \chione and \chitwo. 
In the prior configurations where \chione is nonzero, we recover similar posterior distributions across the \Phenom results, though when \chitwo is forced to zero the distribution shifts to slightly higher values with a median \chione that is \ChiOneDifference larger than in the \ac{LVC} prior case. 
This is because the primary \ac{BH} must now account for all spin effects in the data without a contribution from the secondary. 
The \chione posteriors are also consistent with the Bayes factors reported in Table~\ref{tab:table} in favor of prior configurations where the primary \ac{BH} is spinning: for Models A and C (which are nested since Model C can be obtained by fixing $\chione = 0$ in Model A), the Bayes factor can also be calculated by comparing the prior to the posterior at $\chione=0$~\citep{Chatziioannou2014}. 
As evident in the top panel of Figure~\ref{fig:a1_a2}, the prior at $\chione=0$ is larger than the posterior for both waveform approximants, pointing to a Bayes factor in favor of a spinning primary. 
Estimating Bayes factors from the posterior and prior densities is subject to considerable sampling error when considering the tails of the distributions where there are few samples. 
Nonetheless, for the \Phenom approximant, we find the prior probability density at $\chione = 0$ in Model A to be a factor of $\simeq\,\SDRatioChiOne$ larger than the posterior probability density at $\chione = 0$, which is consistent with the Bayes factor between these two prior configurations ($\simeq\,\BFModelCSimeq$:$1$).

We see larger variation in \chitwo across the prior configurations. 
The standard \ac{LVC} prior recovers a broad, uninformative distribution in \chitwo. 
However, when \chione is forced to zero spin, \chitwo is constrained away from zero in all cases; in these prior configurations, we find \chitwo to be consistent with maximally spinning and have ${\chitwo \gtrsim \ChiTwoNinety}$ at the 90\% credible level \citep{Mandel2020}. 
The \EOB results push to slightly higher secondary spins than the \Phenom results with non-spinning primary configurations, with ${\chitwo\,\gtrsim\,\ChiTwoNinetyModelDEOB}$ at the $90\%$ credible level for Model D-\EOB. 
In all cases where spin misalignment is allowed, we find a preference for some degree of misalignment in the spins; at the $90\%$ credible level, we find ${\thetatwo > \ThetaTwoNinetyModelC}$ in Model C, ${\thetatwo > \ThetaTwoNinetyModelD}$ (${\thetatwo > \ThetaTwoNinetyModelDEOB}$) in Model D (Model D-\EOB), and ${\thetatwo > \ThetaTwoNinetyModelE}$ in Model E. 
Thus, we find that precession (albeit possibly immeasurable) is permitted in all prior configurations that allow for spin tilts.

\section{Discussion and Conclusions}\label{sec:discussion}

GW190412 is an astrophysically compelling event that resides in a previously unobserved region of \ac{BBH} parameter space.  
The effective inspiral spin of the system indicates that at least one of the component \acp{BH} is spinning. 
This work investigates whether a spinning primary \ac{BH} or a spinning secondary \ac{BH} is better supported by the data, and how these hypotheses affect the inferred parameters of GW190412. 

Our main results are summarized in Table~\ref{tab:table}. 
The broad \ac{LVC} prior (Model A) with both \acp{BH} spinning is preferred over the other prior configurations, despite the larger prior volume. 
The degree of preference depends on the waveform approximant used, as the effect of waveform systematics are nonnegligible for this event~\citep{GW190412}. 
We recover marginal support in favor of Model A compared to the prior configuration where only the primary is spinning (Model B). 
When using the \Phenom approximant we find marginal to moderate evidence in favor of Model A compared to prior configurations where only the secondary is allowed to spin (Models C--F), whereas with the \EOB approximant we find strong evidence in support of the \ac{LVC} prior configuration compared to priors with a non-spinning primary. 
The data strongly support Model A over the hypothesis where neither \ac{BH} is spinning (Model G) for both waveform approximants. 

The \Phenom approximant gives broader parameter constraints than the \EOB approximant in both \citet{GW190412} and this work.
In Figure~\ref{fig:corner}, we see that the non-spinning primary prior configurations move the posterior distributions for \q (\chieff) to higher (lower) values to better allow the secondary to account for the spin information in the signal. 
This comes at the cost of matching the data, as the maximum likelihood values are $\LikelihoodDifferencePhenomDex~\mathrm{dex}$ lower for prior configurations where only the secondary is spinning compared to Model A. 
Whereas the \Phenom approximant measures \q (\chieff) to be ${\lesssim \PhenomMassRatioHigh}$ (${\lesssim \PhenomEffectiveSpinHigh}$) and the $90\%$ credible level, the \EOB approximant recovers ${\lesssim \EOBMassRatioHigh}$ (${\lesssim \EOBEffectiveSpinHigh}$). 
The lower mass ratio and higher effective spin from the \EOB analysis makes it more difficult for the data to accommodate a non-spinning primary, with maximum likelihood values that are ${\LikelihoodDifferenceEOBDex~\mathrm{dex}}$ lower for the prior configuration where only the secondary is spinning. 
Despite the larger prior volume, we find a spinning primary hypothesis to be favored over a non-spinning primary hypothesis by a Bayes factor of $\gtrsim\BFModelDEOBApprox$:$1$. 

The prospect of a non-spinning primary \ac{BH} was explored in \citet{Mandel2020}.
\citet{Mandel2020} reweighted the publicly released \chieff posterior samples from the \EOB analysis~\citep{GW190412} in order to apply a prior that assumes a non-spinning primary and a secondary that has a spin aligned with the orbital angular momentum. 
This approach assumes that there is a single measurable spin degree of freedom from GW190412 that is identified with \chieff, and that there is no information about spin precession. 
We instead reanalyze the data under the desired prior, thus imposing no such restrictions about how spins are measured. 
Our analysis results in similar constraints on the secondary spin (Model D with \EOB) as \citet{Mandel2020}, but a different estimate of the mass ratio; we find ${\qLowEOBModelD \leq \q \leq \qHighEOBModelD}$ at the $90\%$ level, compared to ${0.27 \leq \q \leq 0.36}$.
This difference could be attributed to the assumptions of \citet{Mandel2020} about spin measurability. For example, we find 
that the data contain small (but nonnegligible) information about spin precession. 
Additionally, the leading-order spin
term in the \ac{GW} phase is not identical to \chieff~\citep{Poisson1995}, with the difference between the two being more
prominent for unequal mass systems such as GW190412. 
This suggests that the relation between \chitwo and $q$ cannot be fully
explored when considering only \chieff. 
Both \citet{Mandel2020} and this study conclude that the assumption of a non-spinning primary requires a highly spinning secondary, although we find that the corresponding Bayes factors disfavor this
scenario.

Regardless of our prior assumptions, we find the positive effective spin and unequal masses of GW190412 to be robust conclusions. 
However, we do see a shift in the posterior distributions across our prior assumptions. 
With only the secondary \ac{BH} spinning, we recover higher values for \q and lower values for \chieff and \chip. 
The component spins are affected more dramatically; forcing a non-spinning primary causes the secondary's spin magnitude posterior to significantly increase and rail against the physical boundary at $\chitwo=1$. 

The sensitivity of parameter-estimation results to the choice of prior highlights the importance of choosing an appropriate prior when interpreting observations. 
One will never find a spinning primary \ac{BH} if spins are always restricted to be zero; astrophysical models are uncertain and need to be constrained by observations. 
To this end, we can construct prior distributions using a population of observations. 
Performing hierarchical inference enables inference of both individual event's properties and those of the population~\citep{Mandel2010,O2RandP,Galaudage2019}, in effect using the set of observations to construct an empirical prior. 
These inferences may use a branching fraction to consider models from different formation channels~\citep{Stevenson2017,Talbot2017,Vitale2017a,Zevin2017b} or use a phenomenological model to describe the underlying population~\citep{Fishbach2018,Roulet2019,Wysocki2019,Fishbach2020}; they may even encode prior odds for different channels~\citep{Kimball2020}. 
Using wide, uninformative priors, as done by the \ac{LVC}, enables parameter-estimation results to be reweighted by different priors, as required for a hierarchical population analysis~\citep{Mandel2019,Thrane2019}. 

Both the moderately spinning primary and highly spinning secondary interpretations for GW190412 provide unprecedented constraints on astrophysical formation scenarios. 
If GW190412 is the product of isolated binary evolution, our results indicate that the paradigm of negligible natal spin for the first-born \ac{BH} in \ac{BBH} merger progenitors may need to be revised~\citep{Kushnir2016,Zaldarriaga2018,Fuller2019b}. 
Recent work has shown that if post-main-sequence angular momentum transport is not too strong, the first-born \ac{BH} in \ac{BBH} progenitors can be highly spinning from either a Case A (main sequence) mass transfer episode or post-main-sequence tidal spin-up~\citep{Qin2019}. 
However, it is unclear if these systems will become \acp{BBH} with tight enough orbits to merge within a Hubble time. 
Alternatively, GW190412 could be of dynamical origin, with the primary \ac{BH} being the product of one (or more) \ac{BBH} mergers. 
The canonical dynamical scenario---formation in a classical globular cluster~\citep{Benacquista2013}---also struggles to match the parameters of GW190412. 
To be retained in a globular cluster, the natal spins of first-generation \acp{BH} need to be small~\citep[e.g.,][]{Rodriguez2019}. 
In this case, the merger product of two \acp{BH} will form a second-generation \ac{BH} with a dimensionless spin of $\chi \approx 0.67$: above the measurement of \chione in GW190412, which is ${\PhenomAlow \leq \chione \leq \PhenomAhigh}$ with the \Phenom approximant and ${\EOBAlow \leq \chione \leq \EOBAhigh}$ with the \EOB approximant. 
The second-generation globular cluster scenario for GW190412's primary \ac{BH} is also highly disfavored from phenomenological models of hierarchical mergers, which find an odds ratio of $\gtrsim 1000$:$1$ in favor of a GW190412 being a merger of two first-generation \acp{BH} rather than the merger of a first- and second-generation \ac{BH} in a globular cluster~\citep{Kimball2020}. 
Though globular clusters typically cannot retain higher than second-generation merger products due to the relativistic recoil kicks at merger, nuclear clusters~\citep{Gerosa2020}, \ac{AGN} disks~\citep{Tagawa2020}, and high-metallicity super star clusters~\citep{Rodriguez2020} have all been proposed for the formation of GW190412 analogs via hierarchically merging \acp{BH}. 
Other more exotic channels have also been proposed for forming GW190412, such as GW190412 resulting from a $3+1$ hierarchical quadruple stellar system~\citep{Hamers2020}, though \ac{BBH} merger rates from such channels are highly uncertain. 
Explaining \ac{GW} observations requires astrophysical models that can produce systems with both parameters and event rates that are consistent with the measured values.

While the formation scenario for GW190412 is to be determined, the correct interpretation of GW190412's component spins (and those of future \ac{GW} observations) is paramount for constraining viable formation mechanisms. 
As the \ac{GW} detector network continues its observational campaign~\citep{LVC_ObservingScenarios}, additional observations of asymmetric and spinning systems (or lack thereof) will further inform the astrophysical channels that lead to the formation of merging \acp{BBH}. 

Posterior samples for the parameter estimation of GW190412 using our suite of analyses, as well as model evidences, are available on Zenodo~\citep{GW190412_priors_dataset}.

\acknowledgments
The authors would like to thank Richard Udall and Richard O'Shaughnessy for performing parallel calculations using different pipelines, and Rory Smith for useful comments on this manuscript. 
The authors also thank the anonymous referee for helpful suggestions that improved this Letter. 
This work is supported by the National Science Foundation under grant No.\ PHY-1912648. 
M.Z. acknowledges support from CIERA and Northwestern University. 
C.P.L.B. is supported by the CIERA Board of Visitors Professorship. 
The Flatiron Institute is supported by the Simons Foundation.
S.V. acknowledges support of the MIT physics department through the Solomon Buchsbaum Research Fund, the National Science Foundation, and the LIGO Laboratory. 
LIGO was constructed by the California Institute of Technology and Massachusetts Institute of Technology with funding from the National Science Foundation and operates under cooperative agreement PHY-0757058.
This work used computing resources at CIERA funded by National Science Foundation under grant No.\ PHY-1726951, and the computational resources and staff contributions provided for the Quest high performance computing facility at Northwestern University which is jointly supported by the Office of the Provost, the Office for Research, and Northwestern University Information Technology. 
This research has made use of data obtained from the Gravitational Wave Open Science Center (\href{https://www.gw-openscience.org}{www.gw-openscience.org}; \citealt{GWOSC}), a service of LIGO Laboratory, the LIGO Scientific Collaboration and the Virgo Collaboration.
LIGO is funded by the US National Science Foundation.
Virgo is funded by the French Centre National de Recherche Scientifique (CNRS), the Italian Istituto Nazionale della Fisica Nucleare (INFN) and the Dutch Nikhef, with contributions by Polish and Hungarian institutes.

\software{Bilby~\citep{Ashton2019,Romero-Shaw2020a}, iPython~\citep{ipython}, Matplotlib~\citep{matplotlib}, NumPy~\citep{numpy,numpy2}, Pandas~\citep{pandas}, SciPy~\citep{scipy}}

\clearpage
\bibliography{library}
\bibliographystyle{aasjournal}

\end{document}